\documentclass[12pt,a4paper]{article}

\usepackage{ifthen}
\newboolean{pdflatex}
\setboolean{pdflatex}{true}

\newboolean{articletitles}
\setboolean{articletitles}{true}

\newboolean{uprightparticles}
\setboolean{uprightparticles}{false}

\newboolean{inbibliography}
\setboolean{inbibliography}{false}

\usepackage{dcolumn}
\usepackage{placeins}

\newcommand\Tstrut{\rule{0pt}{2.6ex}}

\newcommand\Bstrut{\rule[-1.2ex]{0pt}{0pt}}

\usepackage{microtype}
\usepackage{lineno}  
\usepackage{xspace}
\usepackage{caption}

\usepackage{graphicx} 
\usepackage{color}
\usepackage{colortbl}
\graphicspath{{./figs/}}

\usepackage{amsmath}
\usepackage{amssymb}
\usepackage{amsfonts}
\usepackage{upgreek}

\newcommand*\patchAmsMathEnvironmentForLineno[1]{%
\expandafter\let\csname old#1\expandafter\endcsname\csname #1\endcsname
\expandafter\let\csname oldend#1\expandafter\endcsname\csname
end#1\endcsname
 \renewenvironment{#1}%
   {\linenomath\csname old#1\endcsname}%
   {\csname oldend#1\endcsname\endlinenomath}%
}
\newcommand*\patchBothAmsMathEnvironmentsForLineno[1]{%
  \patchAmsMathEnvironmentForLineno{#1}%
  \patchAmsMathEnvironmentForLineno{#1*}%
}
\AtBeginDocument{%
\patchBothAmsMathEnvironmentsForLineno{equation}%
\patchBothAmsMathEnvironmentsForLineno{align}%
\patchBothAmsMathEnvironmentsForLineno{flalign}%
\patchBothAmsMathEnvironmentsForLineno{alignat}%
\patchBothAmsMathEnvironmentsForLineno{gather}%
\patchBothAmsMathEnvironmentsForLineno{multline}%
\patchBothAmsMathEnvironmentsForLineno{eqnarray}%
}

\usepackage{hyperref}
\usepackage[all]{hypcap}

\def\lhcb {\mbox{LHCb}\xspace}

\def\babar  {\mbox{BaBar}\xspace}
\def\belle  {\mbox{Belle}\xspace}

\ifthenelse{\boolean{uprightparticles}}%
{

 \def\Pmu         {\ensuremath{\upmu}\xspace}

 \def\Ppi         {\ensuremath{\uppi}\xspace}                 
                  
 \def\Prho        {\ensuremath{\uprho}\xspace}

 \def\Ppsi        {\ensuremath{\uppsi}\xspace}

 \def\PDelta      {\ensuremath{\Delta}\xspace}                 
 \def\PXi      {\ensuremath{\Xi}\xspace}                 
 \def\PLambda      {\ensuremath{\Lambda}\xspace}                 
 \def\PSigma      {\ensuremath{\Sigma}\xspace}                 
 \def\POmega      {\ensuremath{\Omega}\xspace}                 
 \def\PUpsilon      {\ensuremath{\Upsilon}\xspace}

 \def\PB      {\ensuremath{\mathrm{B}}\xspace}                 
                  
 \def\PD      {\ensuremath{\mathrm{D}}\xspace}

 \def\PJ      {\ensuremath{\mathrm{J}}\xspace}                 
 \def\PK      {\ensuremath{\mathrm{K}}\xspace}

 \def\Pb      {\ensuremath{\mathrm{b}}\xspace}                 
 \def\Pc      {\ensuremath{\mathrm{c}}\xspace}

 \def\Pi      {\ensuremath{\mathrm{i}}\xspace}

 \def\Ps      {\ensuremath{\mathrm{s}}\xspace}

}
{

 \def\Pmu         {\ensuremath{\mu}\xspace}

 \def\Ppi         {\ensuremath{\pi}\xspace}                 
                  
 \def\Prho        {\ensuremath{\rho}\xspace}

 \def\Ppsi        {\ensuremath{\psi}\xspace}                 
                  
 \mathchardef\PDelta="7101
 \mathchardef\PXi="7104
 \mathchardef\PLambda="7103
 \mathchardef\PSigma="7106
 \mathchardef\POmega="710A
 \mathchardef\PUpsilon="7107
                  
 \def\PB      {\ensuremath{B}\xspace}                 
                  
 \def\PD      {\ensuremath{D}\xspace}

 \def\PJ      {\ensuremath{J}\xspace}                 
 \def\PK      {\ensuremath{K}\xspace}

 \def\Pb      {\ensuremath{b}\xspace}                 
 \def\Pc      {\ensuremath{c}\xspace}

 \def\Pi      {\ensuremath{i}\xspace}

 \def\Ps      {\ensuremath{s}\xspace}

}

\def\mup        {\ensuremath{\Pmu^+}\xspace}
\def\mun        {\ensuremath{\Pmu^-}\xspace} 
\def\mumu       {\ensuremath{\Pmu^+\Pmu^-}\xspace}

\def\squark    {\ensuremath{\Ps}\xspace}

\def\cquark    {\ensuremath{\Pc}\xspace}

\def\bquark    {\ensuremath{\Pb}\xspace}

\def\pion  {\ensuremath{\Ppi}\xspace}

\def\pip   {\ensuremath{\pion^+}\xspace}
\def\pim   {\ensuremath{\pion^-}\xspace}

\def\rhomeson  {\ensuremath{\Prho}\xspace}
\def\rhoz   {\ensuremath{\rhomeson^0}\xspace}

\def\kaon  {\ensuremath{\PK}\xspace}
  \def\Kbar  {\kern 0.2em\overline{\kern -0.2em \PK}{}\xspace}

\def\Kp    {\ensuremath{\kaon^+}\xspace}
\def\Km    {\ensuremath{\kaon^-}\xspace}

\def\Kstarz  {\ensuremath{\kaon^{*0}}\xspace}
\def\Kstarzb {\ensuremath{\Kbar^{*0}}\xspace}

  \def\Dbar    {\kern 0.2em\overline{\kern -0.2em \PD}{}\xspace}
\def\D       {\ensuremath{\PD}\xspace}

\def\Dz      {\ensuremath{\D^0}\xspace}
\def\Dzb     {\ensuremath{\Dbar^0}\xspace}

\def\Dstarp  {\ensuremath{\D^{*+}}\xspace}

\def\B       {\ensuremath{\PB}\xspace}
\def\Bbar    {\ensuremath{\kern 0.18em\overline{\kern -0.18em \PB}{}}\xspace}

\def\Bz      {\ensuremath{\B^0}\xspace}
\def\Bzb     {\ensuremath{\Bbar^0}\xspace}
\def\Bu      {\ensuremath{\B^+}\xspace}

\def\Bp      {\ensuremath{\Bu}\xspace}

\def\Bs      {\ensuremath{\B^0_\squark}\xspace}

\def\Bdb     {\ensuremath{\Bbar^0}\xspace}

\def\jpsi     {\ensuremath{{\PJ\mskip -3mu/\mskip -2mu\Ppsi\mskip 2mu}}\xspace}
\def\psitwos  {\ensuremath{\Ppsi{(2S)}}\xspace}

  \def\Y#1S{\ensuremath{\PUpsilon{(#1S)}}\xspace}

\def\Lz {\ensuremath{\PLambda}\xspace}
\def\Lbar {\ensuremath{\kern 0.1em\overline{\kern -0.1em\PLambda}}\xspace}

\def\Lb      {\ensuremath{\Lz^0_\bquark}\xspace}

\def\qsq       {\ensuremath{q^2}\xspace}

\def\CP                {\ensuremath{C\!P}\xspace}

\newcommand{\ACP}{\ensuremath{{\cal A}_{\CP}}\xspace}

\def\AT#1     {\ensuremath{A_{\mathrm{T}}^{#1}}\xspace}

\def\C#1      {\ensuremath{\mathcal{C}_{#1}}\xspace}                       
\def\Cp#1     {\ensuremath{\mathcal{C}_{#1}^{'}}\xspace}                    
\def\Ceff#1   {\ensuremath{\mathcal{C}_{#1}^{\mathrm{(eff)}}}\xspace}        
\def\Cpeff#1  {\ensuremath{\mathcal{C}_{#1}^{'\mathrm{(eff)}}}\xspace}     
\def\Ope#1    {\ensuremath{\mathcal{O}_{#1}}\xspace}                      
\def\Opep#1   {\ensuremath{\mathcal{O}_{#1}^{'}}\xspace}

\newcommand{\tev}{\ifthenelse{\boolean{inbibliography}}{\ensuremath{~T\kern -0.05em eV}\xspace}{\ensuremath{\mathrm{\,Te\kern -0.1em V}}\xspace}}
\newcommand{\gev}{\ensuremath{\mathrm{\,Ge\kern -0.1em V}}\xspace}
\newcommand{\mev}{\ensuremath{\mathrm{\,Me\kern -0.1em V}}\xspace}
\newcommand{\kev}{\ensuremath{\mathrm{\,ke\kern -0.1em V}}\xspace}
\newcommand{\ev}{\ensuremath{\mathrm{\,e\kern -0.1em V}}\xspace}
\newcommand{\gevc}{\ensuremath{{\mathrm{\,Ge\kern -0.1em V\!/}c}}\xspace}
\newcommand{\mevc}{\ensuremath{{\mathrm{\,Me\kern -0.1em V\!/}c}}\xspace}
\newcommand{\gevcc}{\ensuremath{{\mathrm{\,Ge\kern -0.1em V\!/}c^2}}\xspace}
\newcommand{\gevgevcccc}{\ensuremath{{\mathrm{\,Ge\kern -0.1em V^2\!/}c^4}}\xspace}
\newcommand{\mevcc}{\ensuremath{{\mathrm{\,Me\kern -0.1em V\!/}c^2}}\xspace}

\def\mum  {\ensuremath{{\,\upmu\rm m}}\xspace}

\def\invfb   {\ensuremath{\mbox{\,fb}^{-1}}\xspace}

\newcommand{\chisq}{\ensuremath{\chi^2}\xspace}

\def\gsim{{~\raise.15em\hbox{$>$}\kern-.85em
          \lower.35em\hbox{$\sim$}~}\xspace}
\def\lsim{{~\raise.15em\hbox{$<$}\kern-.85em
          \lower.35em\hbox{$\sim$}~}\xspace}

\def\ptot       {\mbox{$p$}\xspace}
\def\pt         {\mbox{$p_{\rm T}$}\xspace}

\def\evtgen     {\mbox{\textsc{EvtGen}}\xspace}

\def\geant      {\mbox{\textsc{Geant4}}\xspace}

\def\photos     {\mbox{\textsc{Photos}}\xspace}

\def\pythia     {\mbox{\textsc{Pythia}}\xspace}

\def\tell1  {TELL1\xspace}
\def\ukl1   {UKL1\xspace}

\newcommand{\ie}{\mbox{\itshape i.e.}\xspace}

\newcommand{\BdToKstarmumu}{\ensuremath{{\B^0 \rightarrow K^{*0} \mu^+ \mu^-}}\xspace}
\newcommand{\BdToKstarKpimumu}{\ensuremath{{\B^0 \rightarrow K^{*0} (\rightarrow \Kp\pim) \mu^+ \mu^-}}\xspace}
\newcommand{\BToKmumu}{\ensuremath{{\B^+ \rightarrow K^{+} \mu^+ \mu^-}}\xspace}
\newcommand{\BToKmumubar}{\ensuremath{\B^- \rightarrow K^{-} \mu^+ \mu^-}\xspace}

\newcommand{\BdToKstarmumubar}{\ensuremath{\Bdb \rightarrow \Kstarzb \mu^+ \mu^-}\xspace}
\newcommand{\BdToJpsiKstar}{\ensuremath{{\B^0 \rightarrow J/\psi K^{*0}}}\xspace}
\newcommand{\BdToJpsimumuKstar}{\ensuremath{{\B^0 \rightarrow J/\psi (\rightarrow \mumu) K^{*0}}}\xspace}
\newcommand{\BToJpsiK}{\ensuremath{{\B^+ \rightarrow J/\psi K^{+}}}\xspace}

\newcommand{\BToKstmumu}{\ensuremath{{B \rightarrow K^{(*)} \mu^+ \mu^-}}\xspace}
\newcommand{\BToJpsiKst}{\ensuremath{{B \rightarrow \jpsi K^{(*)}}}\xspace}
\newcommand{\BToKstmumubar}{\ensuremath{\Bbar \rightarrow \Kbar^{(*)} \mu^+ \mu^-}\xspace}

\newcommand{\BsToJpsiKstar}{\ensuremath{\Bs \rightarrow J/\psi \Kstarzb}\xspace}

\newcommand{\Kstmumu}{\ensuremath{K^{(*)} \mup \mun}\xspace}

\newcommand{\AD}{\ensuremath{\mathcal{A}_D}\xspace}
\newcommand{\AP}{\ensuremath{\mathcal{A}_P}\xspace}
\newcommand{\ARAW}{\ensuremath{\mathcal{A}_{\rm raw}}\xspace}

\usepackage{cite}
\usepackage{mciteplus}

\usepackage{longtable}

\begin{document}

\renewcommand{\thefootnote}{\fnsymbol{footnote}}
\setcounter{footnote}{1}

\begin{titlepage}
\pagenumbering{roman}

\vspace*{-1.5cm}
\centerline{\large EUROPEAN ORGANIZATION FOR NUCLEAR RESEARCH (CERN)}
\vspace*{1.0cm}
\hspace*{-0.5cm}
\begin{tabular*}{\linewidth}{lc@{\extracolsep{\fill}}r}
\ifthenelse{\boolean{pdflatex}}
{\vspace*{-2.7cm}\mbox{\!\!\!\includegraphics[width=.14\textwidth]{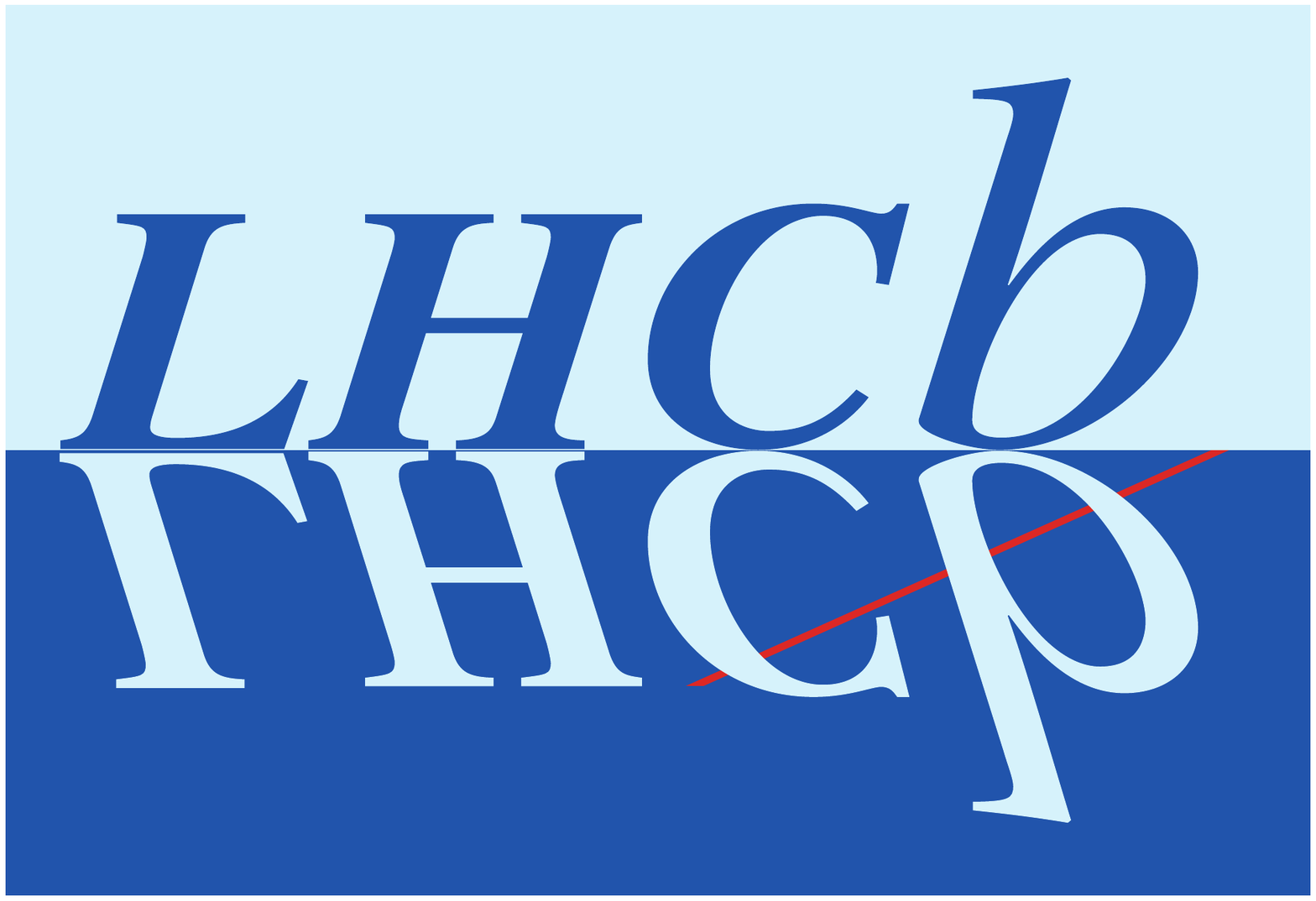}} & &}%
{\vspace*{-1.2cm}\mbox{\!\!\!\includegraphics[width=.12\textwidth]{lhcb-logo.eps}} & &}%
\\
 & & CERN-PH-EP-2014-191 \\ 
 & & LHCb-PAPER-2014-032 \\ 
 & & August 4, 2014
\end{tabular*}
\vspace*{2.0cm}

{\bf\boldmath\huge
\begin{center}
  Measurement of \CP asymmetries in the decays \BdToKstarmumu and \BToKmumu
\end{center}
}

\vspace*{1.5cm}

\begin{center}
The LHCb collaboration\footnote{Authors are listed at the end of this paper.}
\end{center}

\vspace{\fill}

\begin{abstract}
  \noindent
   The direct \CP asymmetries of the decays \BdToKstarmumu and \BToKmumu are measured using $pp$ collision data corresponding to an integrated luminosity of 3.0\invfb collected with the \lhcb detector. The respective control modes \BdToJpsiKstar and \BToJpsiK are used to account for detection and production asymmetries. The measurements are made in several intervals of \mumu invariant mass squared, with the $\phi(1020)$ and charmonium resonance regions excluded. Under the hypothesis of zero \CP asymmetry in the control modes, the average values of the asymmetries are
\begin{align*}
  \ACP(\BdToKstarmumu) &= -0.035 \pm 0.024 \pm 0.003, \\
  \ACP(\BToKmumu) &= \phantom{-}0.012 \pm 0.017 \pm 0.001,
\end{align*}
where the first uncertainties are statistical and the second are due to systematic effects. Both measurements are consistent with the Standard Model prediction of small \CP asymmetry in these decays.
  
\end{abstract}

\vspace*{1.5cm}

\begin{center}
  Published in JHEP 
\end{center}

\vspace{\fill}

{\footnotesize 
\centerline{\copyright~CERN on behalf of the \lhcb collaboration, license \href{http://creativecommons.org/licenses/by/4.0/}{CC-BY-4.0}.}}
\vspace*{2mm}

\end{titlepage}

\newpage
\setcounter{page}{2}
\mbox{~}

\cleardoublepage

\renewcommand{\thefootnote}{\arabic{footnote}}
\setcounter{footnote}{0}

\pagestyle{plain} 
\setcounter{page}{1}
\pagenumbering{arabic}


\section{Introduction}
\label{sec:Introduction}

The processes $\BdToKstarKpimumu$ and $\BToKmumu$ are rare decays\footnote{The inclusion of charge conjugate modes is implied unless explicitly stated.} of $B$ mesons involving $b\rightarrow s$ quark-level transitions, and have small branching fractions, measured as $(1.06 \pm 0.10) \times 10^{-6}$~\cite{LHCb-PAPER-2013-019} and $(4.36 \pm 0.23) \times 10^{-7}$~\cite{LHCb-PAPER-2012-024}. In the Standard Model (SM) there are no tree-level Feynman diagrams for these processes, which proceed via box or electroweak loop (penguin) diagrams. The SM amplitudes are suppressed at loop order, increasing the sensitivity of measurements in these decay channels to physics beyond the SM. Additionally, leading form-factor uncertainties cancel in the measurement of asymmetries, allowing for precise theoretical predictions. Examples include the isospin asymmetry~\cite{LHCb-PAPER-2014-006}, the zero crossing point of the \mumu forward-backward asymmetry~\cite{LHCb-PAPER-2014-007,LHCb-PAPER-2013-019}, and the direct \CP asymmetry, \ACP.

This paper describes measurements of \ACP in \BdToKstarmumu and \BToKmumu decays using data corresponding to 3.0\invfb of integrated luminosity collected by \lhcb in 2011 and 2012, at centre-of-mass energies of 7 and 8\tev, respectively.
The direct \CP asymmetry is defined as
\begin{equation}
\label{eq:ACP}
 \ACP \equiv \frac{\Gamma(\BToKstmumubar) - \Gamma(\BToKstmumu)}{\Gamma(\BToKstmumubar) + \Gamma(\BToKstmumu)},
\end{equation}
where $\Gamma$ is the decay width for the given mode. Non-SM physics contributions could produce interfering diagrams, enhancing the magnitude of \ACP in \BdToKstarmumu decays from the SM prediction of $\mathcal{O}(10^{-3})$~\cite{Altmannshofer:2008dz} to values up to $\pm0.15$~\cite{Alok:2011gv}. Measurements have already been obtained at \lhcb using a data set corresponding to an integrated luminosity of 1.0\invfb, collected in 2011, $\ACP(\BdToKstarmumu) = -0.072 \pm 0.040$~\cite{LHCb-PAPER-2012-021} and $\ACP(\BToKmumu) = 0.000 \pm 0.034$~\cite{LHCb-PAPER-2013-043}, which are the dominant contributions to the world-average values~\cite{PDG2012}. These are consistent both with the SM predictions and with previous results from \babar~\cite{Babar:2012vwa} and \belle~\cite{Wei:2009zv}.

\section{Detector and simulation}
\label{sec:Detector}

The \lhcb detector~\cite{Alves:2008zz} is a single-arm forward
spectrometer covering the \mbox{pseudorapidity} range $2<\eta <5$,
designed for the study of particles containing \bquark or \cquark
quarks. The detector includes a high-precision tracking system
consisting of a silicon-strip vertex detector (VELO) surrounding the $pp$
interaction region, a large-area silicon-strip detector located
upstream of a dipole magnet with a bending power of about
$4{\rm\,Tm}$, and three stations of silicon-strip detectors and straw
drift tubes~\cite{LHCb-DP-2013-003} placed downstream of the magnet. The polarity of the dipole magnet is reversed periodically throughout data-taking.
The tracking system provides a measurement of momentum, \ptot,  with
a relative uncertainty that varies from 0.4\% at low momentum to 0.6\% at 100\gevc.
The minimum distance of a track to a primary $pp$ interaction vertex~(PV), the impact parameter (IP), is measured with a resolution of $(15+29/\pt)\mum$,
where \pt is the component of \ptot transverse to the beam, in~\gevc.
Different types of charged hadrons are distinguished using information
from two ring-imaging Cherenkov (RICH) detectors~\cite{LHCb-DP-2012-003}. Photon, electron and
hadron candidates are identified by a calorimeter system consisting of
scintillating-pad and preshower detectors, an electromagnetic
calorimeter and a hadronic calorimeter. Muons are identified by a
system composed of alternating layers of iron and multiwire
proportional chambers~\cite{LHCb-DP-2012-002}.
The trigger~\cite{LHCb-DP-2012-004} consists of a
hardware stage, based on information from the calorimeter and muon
systems, followed by a software stage, which applies a full event
reconstruction.

Simulated events are used in the process of selecting candidates, examining background contributions, and in determining the efficiency of the selections. In the simulation, $pp$ collisions are generated using
\pythia~\cite{Sjostrand:2006za,*Sjostrand:2007gs} 
 with a specific \lhcb
configuration~\cite{LHCb-PROC-2010-056}.  Decays of hadronic particles
are described by \evtgen~\cite{Lange:2001uf}, in which final-state
radiation is generated using \photos~\cite{Golonka:2005pn}. The
interaction of the generated particles with the detector and its
response are implemented using the \geant
toolkit~\cite{Allison:2006ve, *Agostinelli:2002hh} as described in
Ref.~\cite{LHCb-PROC-2011-006}. The simulated samples are reweighted to model more accurately the data distributions in variables used in the analysis. These include the \pt of the $B$ meson, the number of tracks in the event, and the \chisq of the vertex fit to the final-state tracks, which may differ due to misalignments of the detector and mismodelling of the material description in the VELO region. In addition, information about the IP and momentum resolution is used. The particle identification (PID) performance is corrected to match the data using $\Dstarp \rightarrow (\Dz \rightarrow \Km\pip)\pip$ and $\jpsi \rightarrow \mumu$ control channels.  The \BToKmumu samples are also reweighted for the \pt of the decay products.

\section{Selection of events}
\label{sec:Selection}
Candidates are first required to pass the hardware trigger,
which selects muons with $\pt>1.48\gevc$. In
the subsequent software trigger, at least
one of the final-state particles is required to have both
$\pt>0.8\gevc$ and IP $>100\mum$ with respect to all
of the PVs in the
event. Finally, the tracks of two or more of the final-state
particles are required to form a vertex that is significantly
displaced from the PVs.

All \BdToKstarmumu and \BToKmumu candidates must pass the same initial selection criteria. A requirement on the $B$ candidate vertex fit \chisq per degree of freedom is applied to provide a good quality vertex fit. Additionally, the angle between the momentum vector of the $B$ candidate and the vector between the primary and $B$ candidate decay vertices must be less than 14 mrad, and the $B$ candidates must be consistent with originating from the PV. The tracks from the $B$ candidate decay products are required to be well separated from the PV, helping to reject events where a final-state track does not come from the decay vertex of the $B$ meson. The kaons, pions and muons must be positively identified by PID information from the RICH detectors and muon systems, combined using likelihood functions.

This initial selection is followed by a more stringent selection using multivariate methods, based on a boosted decision tree (BDT)~\cite{Breiman,AdaBoost}. For \BToKmumu decays, simulated signal decays are used for the BDT training, along with data from the upper mass sideband, $5700<m(\Kp\mumu)<6000\mevcc$, which is not used in the remainder of the analysis. The BDT uses a selection of geometric and kinematic variables and has an efficiency of 90\% for signal while removing 95\% of background. Following previous analyses~\cite{LHCb-PAPER-2013-019}, the \BdToKstarmumu BDT training uses a signal sample containing background-subtracted data from the \BdToJpsimumuKstar control mode, and a background sample from the upper mass sideband $5350<m(\Kp\pim\mumu)<7000\mevcc$. This reduces combinatorial background  to small levels.

The \CP asymmetry can vary as a function of the \mumu invariant mass squared, \qsq~\cite{Altmannshofer:2008dz}, and hence the measurement is made in several \qsq bins. The analysis is restricted to \BdToKstarmumu candidates in the range $0.1<\qsq<19.0\gevgevcccc$, and \BToKmumu candidates satisfying $0.1<\qsq<22.0\gevgevcccc$. Three regions are then removed from both samples, corresponding to the $\phi(1020)$ ($0.98<\qsq<1.10\gevgevcccc$), \jpsi ($8.0<\qsq<11.0\gevgevcccc$), and \psitwos ($12.5<\qsq<15.0\gevgevcccc$) resonances. The remaining \BToKstmumu candidates are divided into ${17\,(14)}$ bins that are approximately 1\gevgevcccc wide. The control decay modes, \BToJpsiKst, are selected from the range $8.41<\qsq<10.24\gevgevcccc$. The \BdToKstarmumu candidates are required to have a $\Kp\pim$ mass that lies within $100\mevcc$ of the known \Kstarz mass~\cite{PDG2012}.

Tracks near the edge of the detector acceptance can be swept out by the magnetic field, depending on their charge. This results in the observation of highly asymmetric decay rates for such candidates, as fewer or no candidates with the opposite flavour can be reconstructed. Therefore, fiducial criteria are applied to remove candidates that are reconstructed near the edges of the acceptance. These regions are removed by requiring that the kaon associated with the \BToKstmumu candidates has momentum which satisfies $p_{z} > 2500\,(2000)\mevc$ and $|p_{x}|/(p_{z} - 2500\,(2000))>0.33$, where $p_{z,x}$ are the components of the momentum, measured in \mevc, in the direction of beam travel and in the bending plane, respectively.

There are several background contributions in the signal mass region that require specific vetoes. For the \BToKmumu decays, there is a background from $\Bp \rightarrow \Dzb (\rightarrow \Kp\pim)\pip$ decays where the pions are misidentified as muons. These are removed by computing the mass of the $\Kp\mun$ pair under the $\Kp\pim$ mass hypothesis, and rejecting candidates that satisfy $1850<m(\Kp\pim)<1880\mevcc$. Both modes have backgrounds from \BToJpsiKst events in which a muon from the decay of the \jpsi meson and a final-state hadron are misidentified as each other. These events are vetoed if $m(h^{\pm}\mu^{\mp})$, calculated under the dimuon hypothesis, lies within $60\mevcc$ of the known \jpsi or \psitwos mass, and the hadron can be positively identified as a muon. Other backgrounds for the \BdToKstarmumu decay mode include \BToKmumu events combined with a random pion in the event, $\Lb \rightarrow p\Km\mumu$ decays where at least one hadron is misidentified, and $\Bs \rightarrow \phi\mumu$ events in which a kaon is misidentified as a pion. These are suppressed using a combination of mass and PID requirements similar to those used for the \BToJpsiKst background. The final peaking background comes from \BdToKstarmumu events in which the kaon and pion are misidentified as each other. Since the charge of the kaon identifies the produced meson as either a \Bz or a \Bzb, a misidentification can directly lead to an incorrect asymmetry being measured.

Therefore, the PID information is used to remove events in which the likelihood functions indicate that the reconstructed pion has a higher probability of being a true kaon than the reconstructed kaon. After the vetoes are applied, all of these backgrounds are reduced to less than 1\% of the level of the signal, and are neglected for the rest of the analysis, as are the singly Cabibbo-suppressed backgrounds $\Bz \rightarrow \rhoz\mumu$ and $\Bp \rightarrow \pip\mumu$.

\section{Measurement of direct \CP asymmetries}
\label{sec:Analysis}
Asymmetries in production rate and detection efficiency may bias the measurements and must be accounted for. To first order and for small asymmetries, the raw asymmetry measured, \ARAW, is related to the \CP asymmetry by
\begin{equation}
 \ARAW(\BToKstmumu) = \ACP(\BToKstmumu) + \AP + \AD,
\end{equation}
where any terms from \Bz mixing are neglected, and the production, \AP, and detection, \AD, asymmetries are given by
\begin{equation}
 \AP \equiv \frac{\sigma(\overline{B}) - \sigma(B)}{\sigma(\overline{B}) + \sigma(B)}\qquad \mathrm{and} \qquad
  \AD \equiv  \frac{\epsilon(\overline{f}) - \epsilon(f)}{\epsilon(\overline{f}) + \epsilon(f)},
\end{equation}
where $\sigma$ represents the $B$ meson production cross-section in the \lhcb acceptance, and $\epsilon$ is the detection and reconstruction efficiency for a given final state. The detection asymmetry has two components, one that arises from the different interaction cross-sections of positive and negative particles with the detector material, and another that is due to differences between the left- and right-hand sides of the detector. The latter effect can be reduced by using data collected with both polarities of the magnet, and taking the average. To account for the remaining asymmetries, the control modes \BToJpsiKst are used. These modes have the same particles in the final state and similar kinematic properties to the \BToKstmumu modes, and hence have similar production and detection asymmetries.
Negligible direct CP violation is expected for the control modes, as confirmed by measurements~\cite{PDG2012, LHCb-PAPER-2013-023}. Assuming that the control modes have zero \CP asymmetry, \ACP can be calculated from
\begin{equation}
\label{eq:ACP2}
 \ACP(\BToKstmumu) = \ARAW(\BToKstmumu) - \ARAW(\BToJpsiKst).
\end{equation}
Differences in the production and detection efficiencies of the control and signal modes are considered as sources of systematic uncertainty.

\begin{figure}[tb]
 \centering
\includegraphics[width=\textwidth]{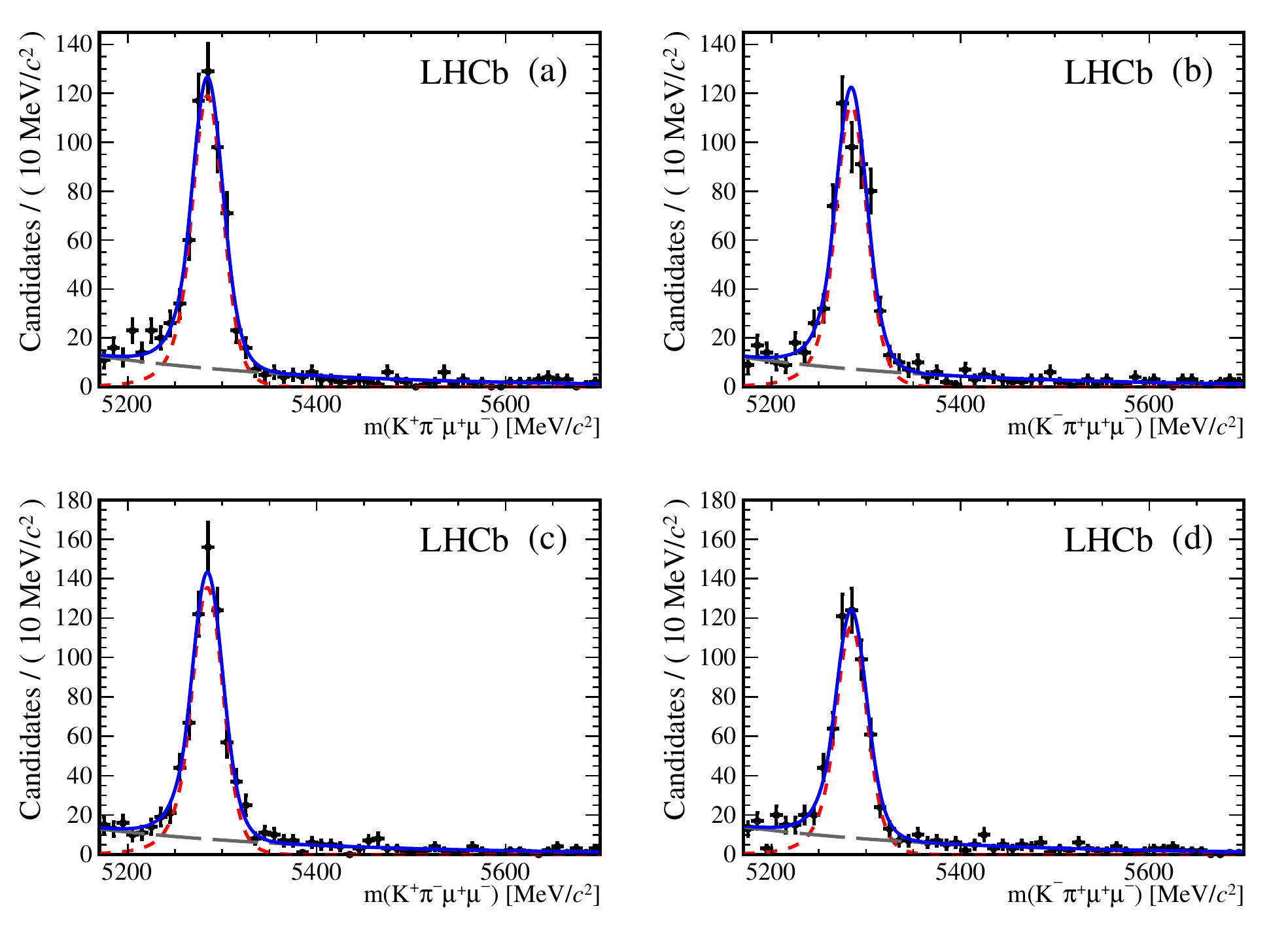}
\caption{Unbinned maximum-likelihood fits to the $\Kp\pim\mumu$ mass distributions of the integrated data set for (a) \BdToKstarmumu and (b) \BdToKstarmumubar decays for one magnet polarity, and (c) \BdToKstarmumu and (d) \BdToKstarmumubar for the other. The blue, solid line represents the total fit, the red, short-dashed line represents the signal component and the grey, long-dashed line represents the combinatorial background.}
 \label{fig:KstarmumuAll}
\end{figure}

\begin{figure}[tb]
 \centering
\includegraphics[width=\textwidth]{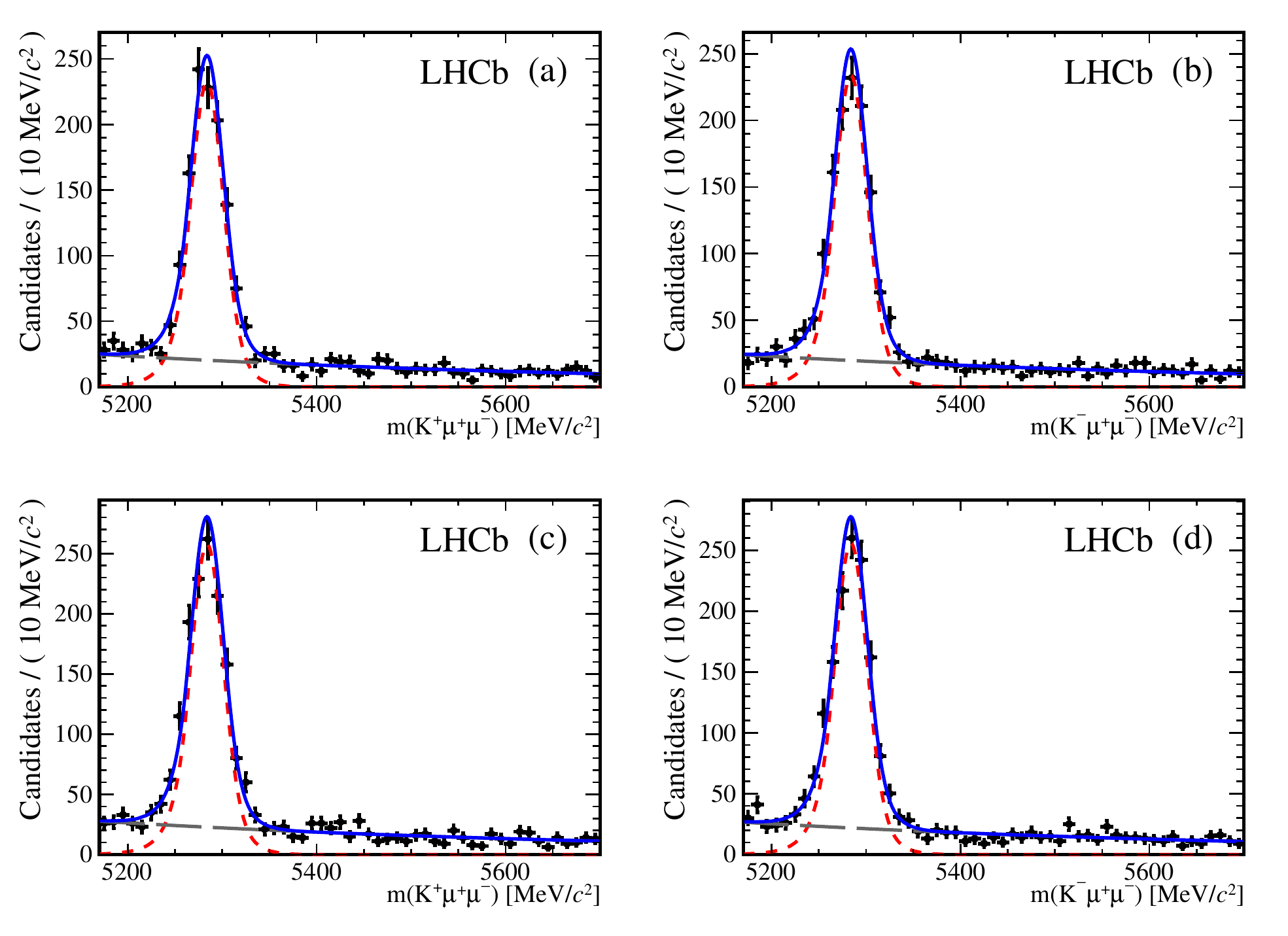}
\caption{Unbinned maximum-likelihood fits to the $\Kp\mumu$ mass distributions of the integrated data set for (a) \BToKmumu and (b) \BToKmumubar decays for one magnet polarity, and (c) \BToKmumu and (d) \BToKmumubar for the other. The blue, solid line represents the total fit, the red, short-dashed line represents the signal component and the grey, long-dashed line represents the combinatorial background.}
 \label{fig:KmumuAll}
\end{figure}

The raw asymmetries are determined via unbinned maximum-likelihood fits to the mass distributions of the candidates. The data set contains approximately 1,000,000 \BToJpsiK, 320,000 \BdToJpsiKstar, 4600 \BToKmumu, and 2200 \BdToKstarmumu signal events in the $B$ mass range $5170<m(\Kstmumu)<5700\mevcc$. The fit shapes used are very similar for all four modes. The signal component is the sum of two Crystal Ball functions~\cite{Skwarnicki:1986xj}, with common mean and tail parameters, but different widths, and the combinatorial background is modelled by an exponential function. The \BdToJpsiKstar mode has an extra contribution arising from \BsToJpsiKstar decays, which is modelled by the same pair of Crystal Ball functions as the signal, but with the mean shifted by the $\Bs-\Bz$ mass difference~\cite{PDG2012}.

All four data sets are split by magnet polarity and charge of the kaon, and the \BToKstmumu data sets are also divided into the ${17\,(14)}$ \qsq bins. The fit is first performed on the four \BToJpsiKst data sets, where the higher number of candidates allows a precise determination of the fit shape parameters. Values for the combined $B$ yield and the raw asymmetry, $\ARAW(\BToJpsiKst)$, for each magnet polarity are determined from the fit. The raw asymmetries in the $\BToJpsiKst$ modes are measured to be $-0.015\,(-0.012)$ for one magnet polarity and $-0.013\,(-0.014)$ for the other. 
The signal shape parameters are then fixed for the fit to the \BToKstmumu mode in each \qsq bin, and the values for the combined $B$ yield and $\ARAW(\BToKstmumu)$ are determined from these fits in the same way. The fits performed on the \BToKstmumu data sets split by kaon charge and magnet polarity are shown in Figs.~\ref{fig:KstarmumuAll} and \ref{fig:KmumuAll}.

The values for $\ACP(\BToKstmumu)$ are determined according to Eq.~\ref{eq:ACP2} for each magnet polarity, and the arithmetic mean of the resulting two values provides the final value for \ACP in each \qsq bin. To obtain an overall value of \ACP across all \qsq bins, an average, weighted by the signal yield and efficiency in each bin, is calculated,
 \begin{equation}
 \ACP = \frac{\sum_{i} (N_{i}\ACP^{i})/\epsilon_{i}}{\sum_{i}N_{i}/\epsilon_{i}},
\end{equation}
where $N_{i}$, $\epsilon_i$, and $\ACP^i$ are the signal yield, signal efficiency, and the value of the \CP asymmetry in the $i\mathrm{th}$ \qsq bin.

\begin{table}[htb]
 \centering
\caption{Summary of the sources of systematic uncertainty for the measurements of $\ACP(\BToKstmumu)$. The ranges shown in parentheses indicate the minimum and maximum values of the systematic uncertainties in different \qsq bins, while the numbers outside the parentheses are the values averaged over \qsq. These may be outside the ranges as the uncertainties are determined by methods affected by statistical fluctuations. There is no systematic uncertainty due to duplicate candidates in the \BToKmumu decay. }
\begin{tabular}{ccc}
Source & \BdToKstarmumu & \BToKmumu \\ \hline
\Tstrut
Kinematic differences & $0.0015\,(0.0025-0.0118)$ & $0.0007\,(0.0007-0.0040)$ \\
Signal shape & $0.0018\,(0.0003-0.0057)$ & $0.0001\,(0.0001-0.0039)$ \\
Background shape & $0.0015\,(0.0016-0.0205)$ & $0.0002\,(0.0001-0.0012)$ \\
Duplicate candidates & $0.0015\,(0.0001-0.0061)$ & $-$\\
\hline
Total & $0.0032\,(0.0063-0.0215)$ & $0.0007\,(0.0011-0.0043)$ \\
    \end{tabular}
  \label{tab:systs}
\end{table}

\section{Systematic uncertainties}
\label{sec:Systematics}
The systematic effects that require consideration are all of a similar magnitude for the \BdToKstarmumu decays, and are listed in order of importance for the \BToKmumu analysis.

In the construction of Eq.~\ref{eq:ACP2}, an assumption is made that the kinematic properties of the particles in the control and signal modes are identical, and therefore the production and detection asymmetries are the same for both modes. However, because the muons from the control mode must originate from the decay of a \jpsi meson, there is a slight difference in the kinematic properties. To estimate the effect that this may have on the result, the data from the control mode are reweighted to the signal mode data so that the distributions match in a chosen kinematic variable. The raw asymmetry, which is approximately the sum of the production and detection asymmetries, is then recalculated from a fit to the weighted data. The difference between the values with the weighted and unweighted data is taken as a contribution to the systematic uncertainty. This procedure is repeated for eight kinematic variables including the momenta and pseudorapidity of the particles and the decay time of the $B$ meson. The sum in quadrature of the differences for each variable is assigned as the overall systematic uncertainty in each \qsq bin. 

In the mass fit, different functions are used to check if the shape used affects the result. The fit is repeated, first replacing the signal component with an Apollonios function, which is the exponential of a hyperbola combined with a low-mass power-law tail~\cite{Santos:2013gra}, and a second time with a second-order Chebychev polynomial modelling the combinatorial background. The differences in the fit results with respect to the nominal fit are assigned as systematic uncertainties.

For the \BdToKstarmumu channel, a further source of systematic uncertainty arises from events that contain duplicate candidates, one with the kaon and pion identified correctly, and one with them swapped. The PID requirement described earlier removes one of each pair of these candidates, but may occasionally select the incorrect candidate. The fit is repeated with both candidates weighted by a factor of one-half, \ie assuming both are equally likely to be correct, rather than with one of them removed. The difference in the fit result is taken as the systematic uncertainty associated with the choice of selection of events with kaon-to-pion swaps. Backgrounds from other decays that are not fully removed by the selection are assumed to exhibit no \CP asymmetry.

None of the systematic uncertainties is larger than 20\% of the statistical uncertainty in any \qsq bin, and the overall systematic uncertainty is less than 7\% of the statistical one. A summary of the systematic uncertainties, indicating the range of each uncertainty across the \qsq bins along with the values for the full data set, is given in Table~\ref{tab:systs}.

\begin{table}[tb]
  \centering
  \caption{Values of \ACP in \BdToKstarmumu decays in each of the 14 \qsq bins used in the analysis. The first uncertainties are statistical and the second are systematic.}
    \begin{tabular}{r@{$-$}l r@{$\,\pm\,$}l c@{$\,\pm\,$}c@{$\,\pm\,$}c}
\multicolumn{2}{c}{ $\qsq$ bin [$\mathrm{GeV}^{2}/{c}^{4}$]} & \multicolumn{2}{c}{Yield} & \multicolumn{3}{c}{\ACP} \Bstrut \\ \hline
\Tstrut
0.10&0.98 & $304$&$18$ & $-$0.087 & 0.060 & 0.006\\ 
1.10&2.00 & $105$&$11$ &$-$0.176 & 0.106 & 0.009\\ 
2.00&3.00 & $120$&$13$ &$-$0.146 & 0.102 & 0.008\\ 
3.00&4.00 & $101$&$12$ &$-$0.013 & 0.113 & 0.014\\ 
4.00&5.00 & $120$&$13$ &$-$0.076 & 0.106 & 0.012\\ 
5.00&6.00 & $143$&$13$ &$-$0.030 & 0.097 & 0.009\\ 
6.00&7.00 & $144$&$14$ &\phantom{+}0.020 & 0.095 & 0.008\\ 
7.00&8.00 & $177$&$15$ &\phantom{+}0.099 & 0.087 & 0.006\\ 
11.0&11.8 & $144$&$14$ &$-$0.021 & 0.093 & 0.007\\ 
11.8&12.5 & $147$&$14$ &\phantom{+}0.031 & 0.093 & 0.022\\ 
15.0&16.0 & $205$&$16$ &$-$0.125 & 0.075 & 0.009\\ 
16.0&17.0 & $216$&$16$ &$-$0.002 & 0.074 & 0.010\\ 
17.0&18.0 & $169$&$14$ &$-$0.059 & 0.085 & 0.009\\ 
18.0&19.0 & $105$&$11$ &$-$0.054 & 0.108 & 0.016\\ 
  \hline
    \end{tabular}
  \label{tab:ACP_Kstarmumu}
  \end{table}

\begin{figure}[tb]
 \centering
\includegraphics[width=0.69\textwidth]{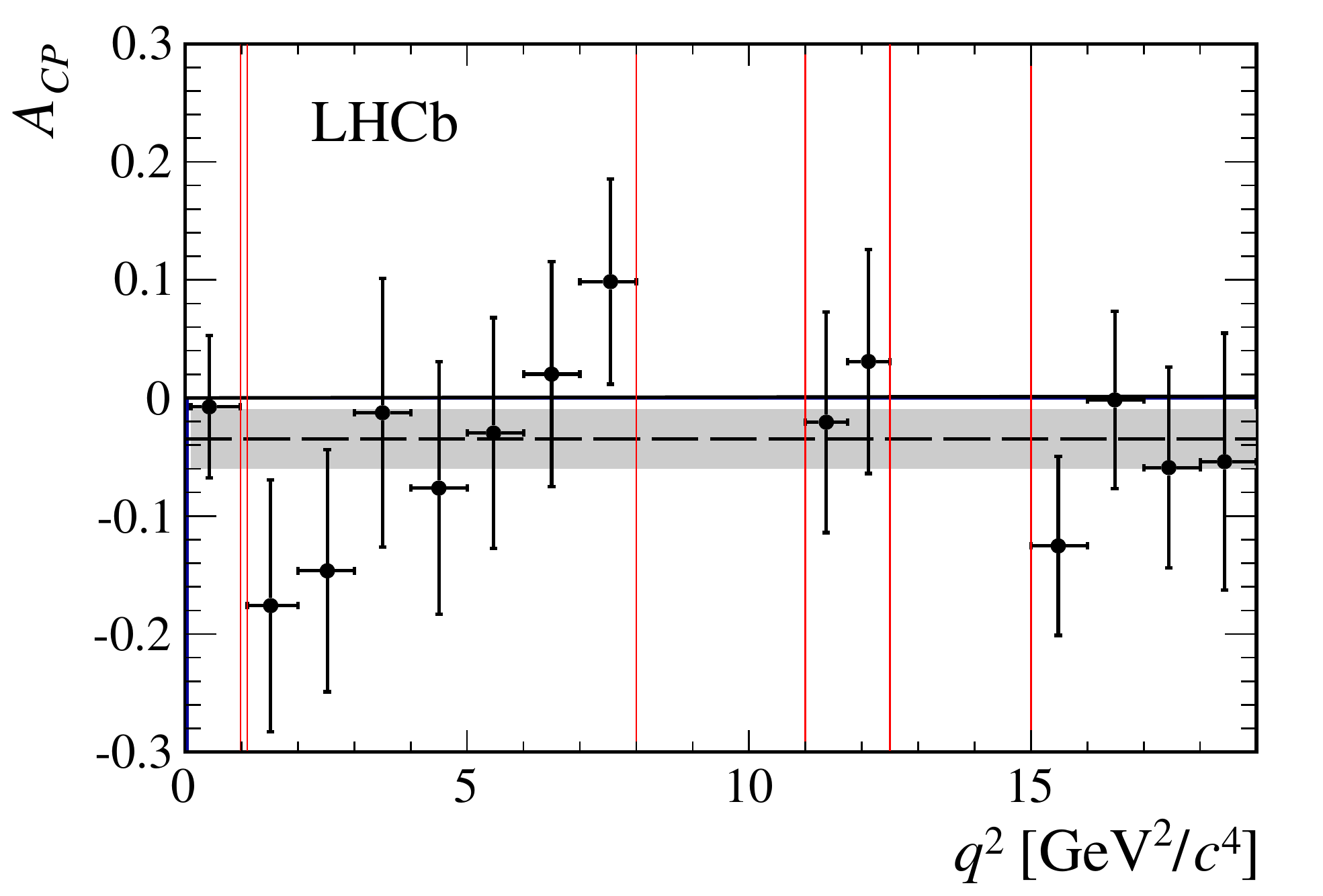}
\caption{Values of \ACP for \BdToKstarmumu decays in each of the the 14 \qsq bins used in the analysis. The error bars are the sum of the statistical and systematic uncertainties in quadrature. The dashed line represents the weighted average value, and the grey band indicates $\pm1\sigma$. The vertical red lines show the $\phi(1020)$, \jpsi, and \psitwos regions, which are vetoed. }
 \label{fig:ACP_Kstarmumu}
\end{figure}


\begin{table}[ht]
  \centering
  \caption{Values of \ACP in \BToKmumu decays in each of the 17 \qsq bins used in the analysis. The first uncertainties are statistical and the second are systematic.}
    \begin{tabular}{r@{$-$}l r@{$\,\pm\,$}l c@{$\,\pm\,$}c@{$\,\pm\,$}c}
\multicolumn{2}{c}{ $\qsq$ bin [$\mathrm{GeV}^{2}/{c}^{4}$]}& \multicolumn{2}{c}{Yield} & \multicolumn{3}{c}{\ACP} \Bstrut \\ \hline
\Tstrut
0.10&0.98 & $387$&$22$& \phantom{+}0.088 & 0.057 & 0.001\\ 
1.10&2.00 & $277$&$19$& $-$0.004 & 0.068 & 0.002\\ 
2.00&3.00 & $367$&$22$& \phantom{+}0.042 & 0.059 & 0.001\\ 
3.00&4.00 & $334$&$21$& $-$0.034 & 0.063 & 0.001\\ 
4.00&5.00 & $307$&$20$& $-$0.021 & 0.064 & 0.001\\ 
5.00&6.00 & $332$&$21$& \phantom{+}0.031 & 0.062 & 0.002\\ 
6.00&7.00 & $355$&$22$& \phantom{+}0.026 & 0.060 & 0.001\\ 
7.00&8.00 & $371$&$22$& \phantom{+}0.041 & 0.059 & 0.002\\ 
11.0&11.8 & $232$&$18$ & $-$0.047 & 0.076 & 0.002\\ 
11.8&12.5 & $247$&$17$& \phantom{+}0.018 & 0.070 & 0.002\\ 
15.0&16.0 & $287$&$19$& \phantom{+}0.120 & 0.065 & 0.004\\ 
16.0&17.0 & $287$&$19$& \phantom{+}0.028 & 0.066 & 0.001\\ 
17.0&18.0 & $349$&$21$& $-$0.030 & 0.058 &0.001 \\ 
18.0&19.0 & $222$&$17$& $-$0.061 & 0.074 & 0.003\\ 
19.0&20.0 & $121$&$13$& $-$0.048 & 0.105 & 0.003\\ 
20.0&21.0 & $95$&$12$& $-$0.012 & 0.120 & 0.003\\ 
21.0&22.0 & $50$&$8$& $-$0.290 & 0.161 & 0.004\\ 
 \hline
    \end{tabular}
  \label{tab:ACP}
  \end{table}

\section{Results}
\label{sec:Results}

\begin{figure}[tb]
 \centering
\includegraphics[width=0.69\textwidth]{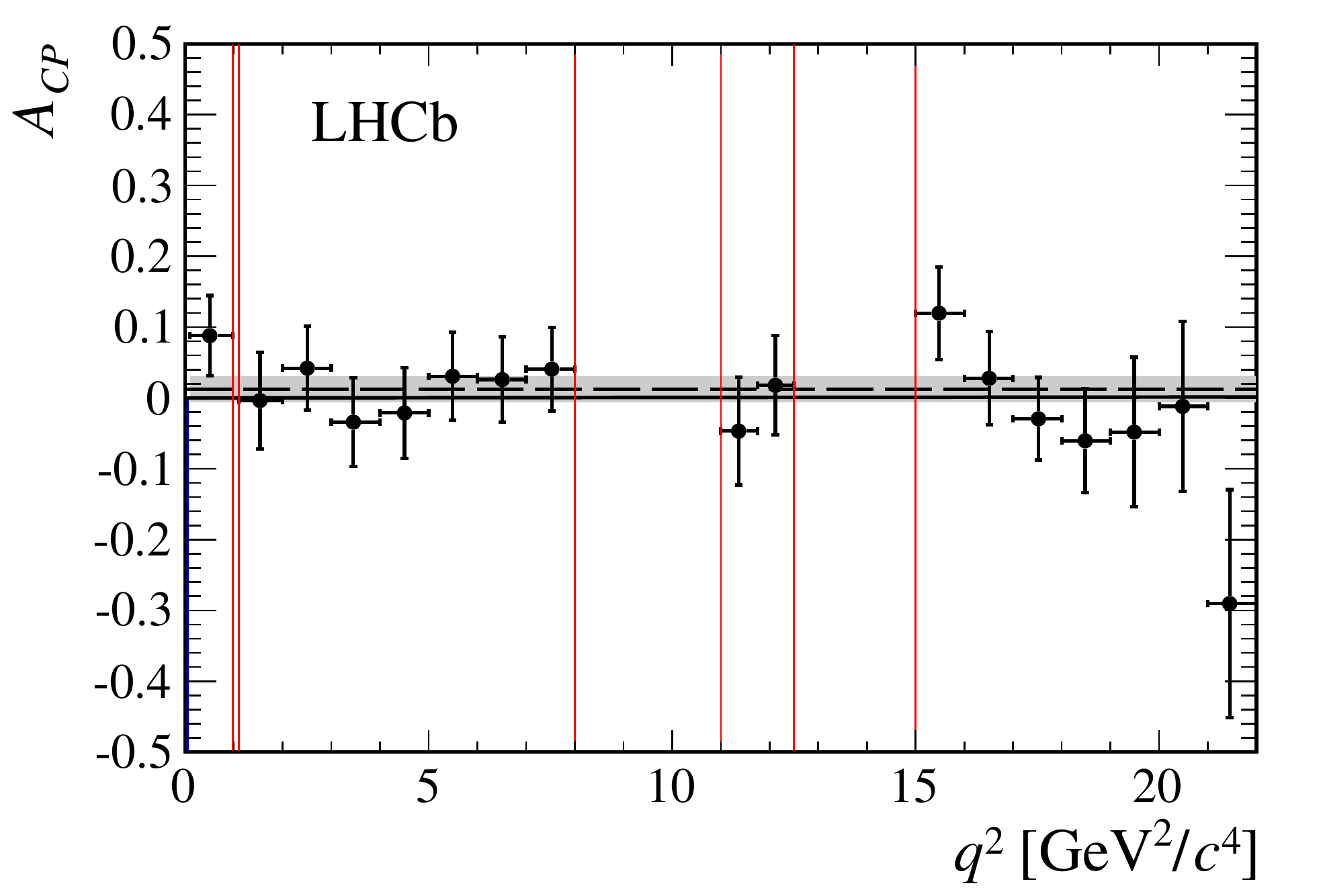}
\caption{Values of \ACP for \BToKmumu decays in each of the the 17 \qsq bins used in the analysis. The error bars are the sum of the statistical and systematic uncertainties in quadrature. The dashed line represents the weighted average value, and the grey band indicates $\pm1\sigma$. The vertical red lines show the $\phi(1020)$, \jpsi, and \psitwos regions, which are vetoed.}
 \label{fig:ACP}
\end{figure}

The results in each \qsq bin are shown in Table~\ref{tab:ACP_Kstarmumu} and Fig.~\ref{fig:ACP_Kstarmumu} for \BdToKstarmumu, and Table~\ref{tab:ACP} and Fig.~\ref{fig:ACP} for \BToKmumu. The values of the \CP asymmetries in \BToKstmumu decays are
\begin{align*}
  \ACP(\BdToKstarmumu) &= -0.035 \pm 0.024 \pm 0.003, \\
  \ACP(\BToKmumu) &= \phantom{-}0.012 \pm 0.017 \pm 0.001,
\end{align*}
where the first uncertainties are statistical and the second are due to systematic effects. They are obtained under the hypothesis of zero \CP asymmetry in the control modes, \BdToJpsiKstar and \BToJpsiK. Both of these results, which supersede the previous 1.0\invfb measurements~\cite{LHCb-PAPER-2012-021,LHCb-PAPER-2013-043}, are consistent with the SM predictions, and the uncertainties on the measurements are almost a factor of two smaller than the previous best values.

\clearpage

\section*{Acknowledgements}

\noindent We express our gratitude to our colleagues in the CERN
accelerator departments for the excellent performance of the LHC. We
thank the technical and administrative staff at the LHCb
institutes. We acknowledge support from CERN and from the national
agencies: CAPES, CNPq, FAPERJ and FINEP (Brazil); NSFC (China);
CNRS/IN2P3 (France); BMBF, DFG, HGF and MPG (Germany); SFI (Ireland); INFN (Italy); 
FOM and NWO (The Netherlands); MNiSW and NCN (Poland); MEN/IFA (Romania); 
MinES and FANO (Russia); MinECo (Spain); SNSF and SER (Switzerland); 
NASU (Ukraine); STFC (United Kingdom); NSF (USA).
The Tier1 computing centres are supported by IN2P3 (France), KIT and BMBF 
(Germany), INFN (Italy), NWO and SURF (The Netherlands), PIC (Spain), GridPP 
(United Kingdom).
We are indebted to the communities behind the multiple open 
source software packages on which we depend. We are also thankful for the 
computing resources and the access to software R\&D tools provided by Yandex LLC (Russia).
Individual groups or members have received support from 
EPLANET, Marie Sk\l{}odowska-Curie Actions and ERC (European Union), 
Conseil g\'{e}n\'{e}ral de Haute-Savoie, Labex ENIGMASS and OCEVU, 
R\'{e}gion Auvergne (France), RFBR (Russia), XuntaGal and GENCAT (Spain), Royal Society and Royal
Commission for the Exhibition of 1851 (United Kingdom).

\addcontentsline{toc}{section}{References}
\setboolean{inbibliography}{true}
\bibliographystyle{LHCb}
\bibliography{main,LHCb-PAPER,LHCb-CONF,LHCb-DP}

\newpage
\centerline{\large\bf LHCb collaboration}
\begin{flushleft}
\small
R.~Aaij$^{41}$, 
B.~Adeva$^{37}$, 
M.~Adinolfi$^{46}$, 
A.~Affolder$^{52}$, 
Z.~Ajaltouni$^{5}$, 
S.~Akar$^{6}$, 
J.~Albrecht$^{9}$, 
F.~Alessio$^{38}$, 
M.~Alexander$^{51}$, 
S.~Ali$^{41}$
% ...

{\footnotesize \it
% ...
$ ^{5}$Clermont Universit\'{e}, Universit\'{e} Blaise Pascal, CNRS/IN2P3, LPC, Clermont-Ferrand, France\\
$ ^{6}$CPPM, Aix-Marseille Universit\'{e}, CNRS/IN2P3, Marseille, France\\
% ...
$ ^{9}$Fakult\"{a}t Physik, Technische Universit\"{a}t Dortmund, Dortmund, Germany\\
% ...
$ ^{37}$Universidad de Santiago de Compostela, Santiago de Compostela, Spain\\
$ ^{38}$European Organization for Nuclear Research (CERN), Geneva, Switzerland\\
% ...
$ ^{41}$Nikhef National Institute for Subatomic Physics, Amsterdam, The Netherlands\\
% ...
$ ^{46}$H.H. Wills Physics Laboratory, University of Bristol, Bristol, United Kingdom\\
% ...
$ ^{51}$School of Physics and Astronomy, University of Glasgow, Glasgow, United Kingdom\\
$ ^{52}$Oliver Lodge Laboratory, University of Liverpool, Liverpool, United Kingdom\\
% ...
}
\end{flushleft}

\end{document}